\documentclass[a4paper,fleqn]{cas-sc}
\usepackage{diagbox}
\usepackage[authoryear,sort&compress]{natbib}
\usepackage{wrapfig}
\usepackage[ruled,vlined]{algorithm2e}
\usepackage{algpseudocode} 
\usepackage{graphicx}
\usepackage{caption}
\usepackage{subcaption}

\def\tsc#1{\csdef{#1}{\textsc{\lowercase{#1}}\xspace}}
\tsc{WGM}
\tsc{QE}
\tsc{EP}
\tsc{PMS}
\tsc{BEC}
\tsc{DE}
\begin{document}
\let\WriteBookmarks\relax
\def\floatpagepagefraction{1}
\def\textpagefraction{.001}
%\shorttitle{EEG-VSWM}
%\shortauthors{Sharma \& Uddin}
%\begin{frontmatter}

\title  [mode = title]{Retracted Citations and Self-citations in Retracted Publications: A Comparative Study of Plagiarism and Fake Peer Review}
%\tnotemark[1,2]

\author[1,2]{Kiran Sharma}\corref{cor1}
\ead{kiran.sharma@bmu.edu.in}
\author[3]{Parul Khurana}
% %\credit{Conceived and designed the analysis; Collected the data;  Performed the analysis; Writing - original draft}
% %
% %
% \author[1,2]{Ziya Uddin}\corref{cor2}
% \ead{ziya.uddin@bmu.edu.in}
% %\credit{Conceived and designed the analysis; Performed the analysis; Critical input}
% %

\address{School of Engineering \& Technology, BML Munjal University, Gurugram, Haryana-122413, India }
\address{Center for Advanced Data and Computational Science, BML Munjal University, Gurugram, Haryana-122413, India }
\cortext[cor1]{Corresponding author}
\address{School of Computer Applications, Lovely Professional University, Phagwara, Punjab-144411, India }
% \cortext[cor2]{Corresponding author2}

%%%%%%%%%%%%%%%%%%%%%%%%%%%%%%%%%%%%%%%%%%%%%%%%%%%%%%%%%%%%%%%%%%%%%%%%%%%%%

%==================================================================
\begin{abstract}
Retracted citations remain a significant concern in academia as they perpetuate misinformation and compromise the integrity of scientific literature despite their invalidation. To analyze the impact of retracted citations, we focused on two retraction categories: plagiarism and fake peer review. The data set was sourced from Scopus and the reasons for the retraction were mapped using the Retraction Watch database. The retraction trend shows a steady average growth in plagiarism cases of 1.2 times, while the fake peer review exhibits a fluctuating pattern with an average growth of 5.5 times. Although fewer papers are retracted in the plagiarism category compared to fake peer reviews, plagiarism-related papers receive 2.5 times more citations. Furthermore, the total number of retracted citations for plagiarized papers is 1.8 times higher than that for fake peer review papers. Within the plagiarism category, 46\% of the retracted citations are due to plagiarism, while 53.6\% of the retracted citations in the fake peer review category are attributed to the fake peer review. The results also suggest that fake peer review cases are identified and retracted more rapidly than plagiarism cases. Finally, self-citations constitute a small percentage of citations to retracted papers but are notably higher among citations that are later retracted in both the categories.
\end{abstract}
%==================================================================

%%Research highlights
%\begin{highlights}
%\item Research highlight 1
%\item Research highlight 2
%\end{highlights}
%==================================================================
\maketitle
%==================================================================

%\end{frontmatter}

%\tableofcontents

%% \linenumbers

%% main text
%==================================================================
\section{Introduction}

Retracted citations refer to citations made to academic papers that have been officially retracted by publishers or journals due to issues such as errors, misconduct, plagiarism, falsified data, or ethical violations. Despite being retracted, these papers often continue to be cited in new research, sometimes without acknowledgment of their retracted status~\citep{Silva2016Why, Gray2019Why, DaSilva2020Reasons}.
The issue of retracted citations poses a serious challenge to the academic community, as retracted papers often continue to be cited despite their invalidation. This practice can spread misinformation and undermine the credibility and integrity of the scientific literature.

The number of citations for retracted articles has increased over time, with a constant increase in the percentage of acknowledging their retraction~\citep{heibi2021qualitative, sharma2024over}. Most of the retracted articles, particularly those published in Nature, Science, and Cell, continue to be cited even after their retraction~\citep{Wang2022Expert-recommended}. \cite{Tang2023Some} study also highlighted that post-retraction citations in the top ranked journals, Nature and Science, account for 47.7\% and 40.9\% of total citations, respectively, with factors such as misconduct, validity issues, and background citation noise contributing to these retractions.

Post-retraction citations are an avoidable phenomenon. Although, retraction decreased the frequency of citation by about 60\%, compared to non-retracted papers, but retracted papers often live on~\citep{Kuhberger2022Self-correction}. Previous research on retracted articles has revealed that, despite being flagged, such studies are still frequently cited as valid across various disciplines~\citep{Bar-Ilan2017Post, sharma2021team}.

In the study by \cite{DeCassai2022Inappropriate} in anesthesiology and intensive care medicine, they examined that 46\% of the articles retracted were cited at least once after retraction, and many authors were unaware of the retraction. \cite{Bolboaca2019Post} investigates the trends and citation patterns that occur after retraction of articles in the field of radiology imaging diagnostics. Post-retraction citations in radiology imaging diagnostic methods are higher than before retraction in 30 out of 54 cases, plagiarism being the most common reason for retraction (31\%). The persistence of post-retraction citations in radiology-imaging diagnostic methods, as well as in other medical fields like radiation oncology, points to a systemic issue in academic publishing. 

Retracted biomedical research papers continue to be cited at relatively high rates, despite the retraction process~\citep{Hagberg2020The}. \cite{hamilton2019continued} also quantified the number and explored the nature of the citations of articles retracted in the radiation oncology literature that occur after publication of the retraction note. The study found that 92\% of the 358 post-retraction citations examined referenced retracted articles as legitimate work. The results of the study emphasize the need for investigators to adhere to good research practices to mitigate the influence and propagation of flawed and unethical research.
\cite{Schneider2020Continued} also presented a case study of long-term post-retraction citation to falsified clinical trial data. They investigated that even 11 years after its retraction, the paper is still being cited positively and uncritically to support a medical nutrition intervention, with no acknowledgment of its 2008 retraction for data falsification.

In addition, \cite{Palla2023Systematic} studied that the number of articles retracted by Indian researchers increased from 2001 to 2020. The main reason for the retraction was duplication and plagiarism. They analyzed that 90\% of the articles retracted by Indian researchers were cited even after the retraction process, with a total decline of 8\% in citations after the retraction process. The protocol proposed by \cite{heibi2022protocol} can be used as a comprehensive framework to analyze the citation patterns of retracted articles. This is due to the importance of increasing awareness and better management of the retraction information. Understanding such patterns can therefore help mitigate the impact of retracted articles on the scientific literature and ensure academic research integrity.
%%~~~~~~~~~~~~~~~~~~~~~~~~~~~~~~~~~~~~~~~~~~~~~~~~~~~~~~~~~~~~~~~~~~~~~~~~~~~~~~~~~~~~~~~~~~~~~~~
\subsection{Research Gap}

Earlier studies have focused mainly on the increase in citations of retracted publications in various fields. \cite{Kocyigit2023Analysis} analyzed retracted articles in the medical literature due to ethical issues. \cite{Qi2016Characteristics} studied the retraction due to fake peer reviews, where publishing journals and authors are concerned. \cite{Kamali2020Plagiarism} studied Iran-associated scientific papers retracted for duplication, plagiarism, and fake peer reviews, calling for immediate intervention and education in ethical research. \cite{wang2025empirical} studied retractions due to honest errors in team size.
On the other hand, \cite{Rivera2018Fake} highlighted that inappropriate authorship and fake peer review are real evils that contribute to lower quality publications. \cite{Bell2022Scholarly} also highlighted that fake peer reviews in scholarly publications are a growing concern, highlighting the need to distinguish genuine reviews and to defend the boundaries between science and society. All of these papers examined various reasons for retractions in the context of the growing number of retracted publications. However, none focused on retracted citations, their subsequent reasons, or self-citations. This study addresses this gap by specifically analyzing the retraction categories of plagiarism and fake peer review, their associated retracted citations, and the reasons behind these retractions. Additionally, it delves deeper into the self-citations reported within retracted citations in both categories.

%==================================================================
\section{Research Objectives}
The study seeks to analyze and investigate the following objectives within the categories of plagiarism and fake peer review retractions:
\begin{enumerate}
\item Trends in publication and retraction for both categories. 
\item Distribution of retracted citations across both categories and the reason for retractions. 
\item Distribution of self-citations within both categories. 
\end{enumerate}
%%~~~~~~~~~~~~~~~~~~~~~~~~~~~~~~~~~~~~~~~~~~~~~~~
\subsection{Research Questions}
The following research questions are designed to guide the investigation of the objectives related to plagiarism and fake peer review retractions:
\begin{itemize}
    \item \textbf{R1:} How have the retraction rates evolved over time in both categories? 
    \item \textbf{R2:} What is the average time of retraction (in years) for plagiarism and fake peer review? 
    \item \textbf{R3:} What is the distribution of retracted citations in plagiarism and fake peer review retractions? 
    \item \textbf{R4:} What share of retracted citations falls under the same retracted category? 
  %  \item \textbf{R5:} What is the distribution of retracted citations before and after retraction of the main paper?
    \item \textbf{R5:} How do self-citations contribute to the total number of citations in plagiarized and fake peer review retracted articles?
    \item \textbf{R6:} Are self-citations more prevalent in one category (plagiarism or fake peer review retraction) than in the other?
%    \item \textbf{R8:} What is the distribution of self-citations before and after retraction of the main paper?
\end{itemize}

%==================================================================

\section{Methodology}

\subsection{Data Description}
\begin{figure}[!ht]
    \centering 
\includegraphics[width=0.7\linewidth]{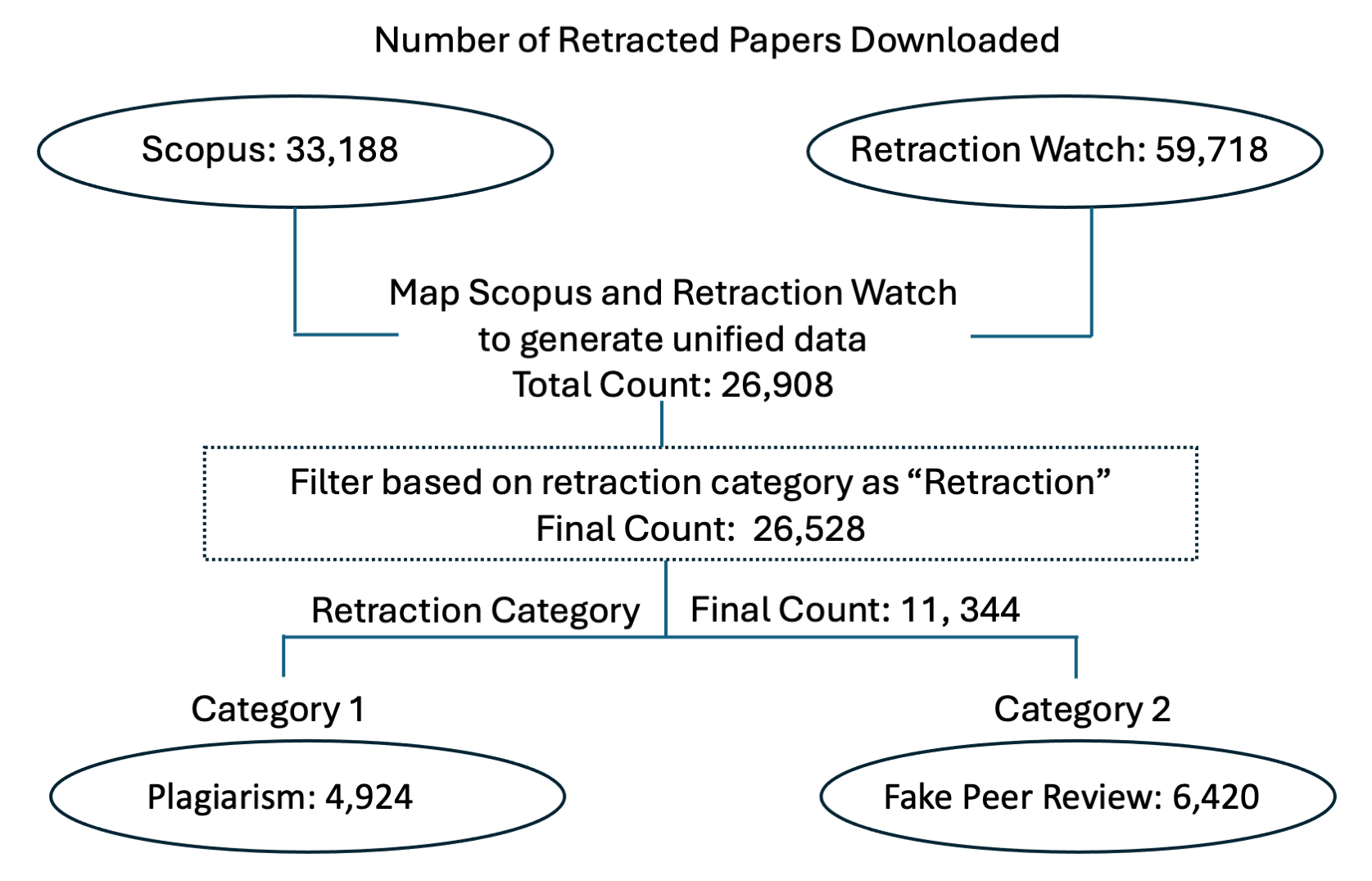} 
\caption{Data flowchart.}
\label{fig:flowchart}
\end{figure}

The study utilized Scopus-sourced scholar publishing data, downloaded on 7 December 2024, comprising a total of 33,188 publications with document type retracted. The Scopus query to extract the retracted publication was: ``\textit{DOCTYPE ( tb )}''. To ensure accurate linkage with retraction records, the dataset was filtered to include only documents with a DOI \cite{khurana2022bibliometric}, resulting in 32,861 entries. These DOIs were then matched with the Retraction Watch database (\url{https://www.crossref.org/blog/news-crossref-and-retraction-watch/}) to identify documents flagged for retraction. This mapping process was successful for 26,908 documents. Subsequently, a filtration step was applied to isolate cases where the nature of the retraction explicitly indicated ``Retraction,'', which produced a data set of 26,528 documents for analysis. Figure~\ref{fig:flowchart} shows the description of the data.

\subsection{Retraction Categories}
We classified the data into two categories: plagiarism and fake peer review, discarding the remaining data. These categories were chosen because of the clear documentation of retraction reasons and the significant number of retractions within them. The plagiarism category includes papers retracted for retraction reasons such as ``\textit{Plagiarism of articles, data, images and texts + Duplication of article, data, image and text}'' and the category fake peer review includes papers retracted for retraction reasons such as ``\textit{Fake peer review + Concerns / Issues with peer review}'' (\url{https://retractionwatch.com/retraction-watch-database-user-guide/retraction-watch-database-user-guide-appendix-b-reasons/}). 
Out of 26,528 filtered documents, we further categorized the data into two retraction categories such as Plagiarism - category 1 and Fake peer review - category 2. We filter 4,924 classified as plagiarism and 6,420 as fake peer review. A total of 156 papers with 1,954 citations appeared with both retraction reasons; hence for simplicity, we excluded these papers. Table~\ref{tab:RetCount} shows the final count of retractions.
 %~~~~~~~~~~~~~~~~~~~~~~~~~~~~~~~~~~~~~~~~~~~~~~~~~~~~~~~~~~~~~~~~~~~~~
%~~~~~~~~~~~~~~~~~~~~~~~~~~~~~~~~~~~~~~~~~~~~~~~~~~~~~~~~~~~~~~~~~~~~~~~

%%~~~~~~~~~~~~~~~~~~~~~~~~~~~~~~~~~~~~~~~~~~~~~~~
\begin{table}[!ht]
\centering
\caption{Retraction count along with citations for both categories.}
\begin{tabular}{|l|c|cc|c|}
\hline
\multicolumn{1}{|c|}{\multirow{2}{*}{\textbf{Retraction Category}}} & \multirow{2}{*}{\textbf{Total Retractions}} & \multicolumn{2}{c|}{\textbf{Number of Retractions}}              & \multirow{2}{*}{\textbf{Total Citations}} \\ \cline{3-4}\multicolumn{1}{|c|}{} &   & \multicolumn{1}{c|}{\textbf{With Citations}} & \textbf{Without Citation}s &                                  \\ \hline
Plagiarism                                                 & 4,924                               & \multicolumn{1}{c|}{4,482}           & 442               & 1,41,891                           \\ \hline
Fake Peer Review                                           & 6,420                               & \multicolumn{1}{c|}{5,197}           & 1,223              & 55,272                            \\ \hline
Total                                                      & 11,344                              & \multicolumn{1}{c|}{9,679}           & 1,665              & 1,97,163                           \\ \hline
\end{tabular}
\label{tab:RetCount}
\end{table}
%%~~~~~~~~~~~~~~~~~~~~~~~~~~~~~~~~~~~~~~~~~~~~~~~
%==================================================================
\section{Results and Discussion}

\subsection{Publication and Retraction Trend}

\begin{figure}[!ht]
    \centering 
\includegraphics[width=0.48\linewidth]{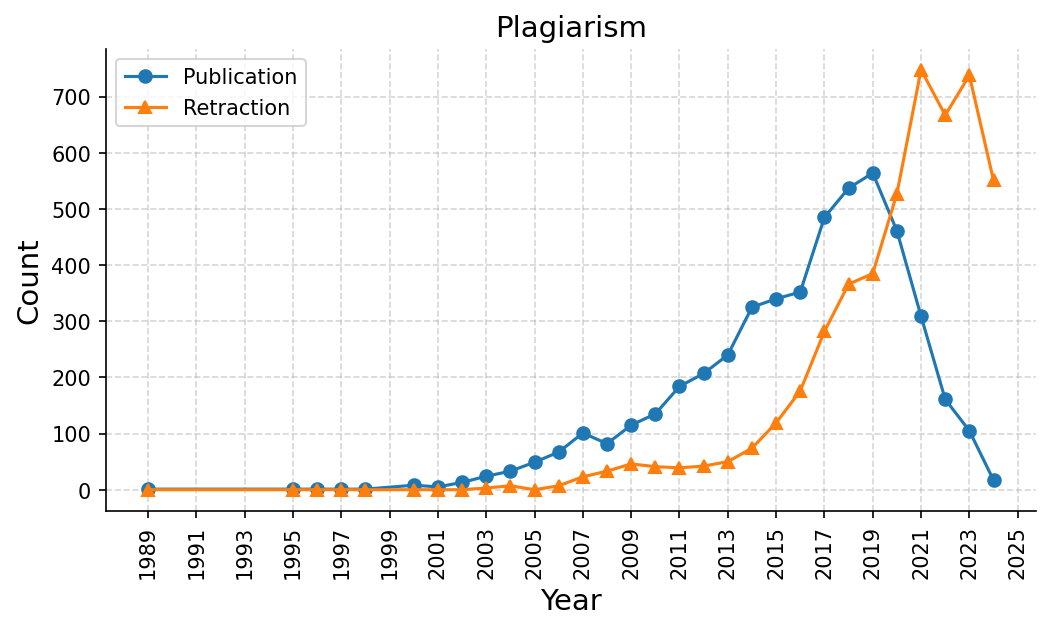} 
\includegraphics[width=0.48\linewidth]{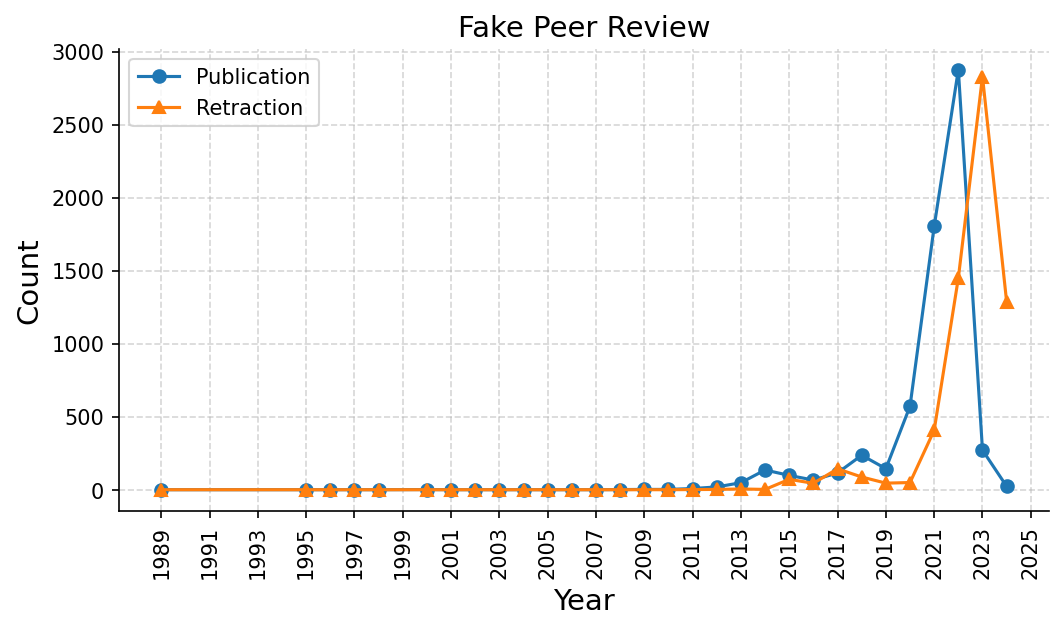} 
\caption{Number of papers published and later got retracted over years. (a) Plagiarism and (b) Fake peer review.}
\label{fig:Trend}
\end{figure}

The number of retracted articles has increased significantly over recent years, driven by various factors~\citep{Steen2013Why, sharma2024over}.
Figure~\ref{fig:Trend}(a) represents the trend of publications and retractions over the years due to plagiarism. Publications show a steady rise from 2005, peaking in 2020, and then declining. The decline after 2020 might reflect stricter plagiarism detection. The overlap in publication and retraction trends suggests a robust but delayed system to identify plagiarism. Retractions increase from 2010, peak in 2021, and then decrease gradually, mirroring the publication trend. The retractions have shown a steady increase over the years, with an average growth rate of 1.2 times.

Fake peer-review is a growing issue in academia~\citep{Hadi2016Fake}. Figure~\ref{fig:Trend}(b) represents the trend of publications and retractions over the years due to fake peer review cases. 
Publications start increasing significantly around 2011 and peak in 2021, suggesting a rise in published works associated with fake peer review.
Retractions follow a similar trend, but lag slightly, with a dramatic increase from around 2017, peaking in 2022, and then sharply declining. In 2021, the retraction rate increased by 8.3 times compared to 2020. In 2022, it rose to 3.6 times the rate of 2021. By 2023, the increase was 1.9 times, indicating a decline in the growth rate of retractions in recent years.
%~~~~~~~~~~~~~~~~~~~~~~~~~~~~~~~~~~~~~~~~~~~~~~~~~~~~~~~~~~~~~~~~~~~~~
%~~~~~~~~~~~~~~~~~~~~~~~~~~~~~~~~~~~~~~~~~~~~~~~~~~~~~~~~~~~~~~~~~~~~~~~

\begin{figure}[!ht]
    \centering 
\includegraphics[width=0.7\linewidth]{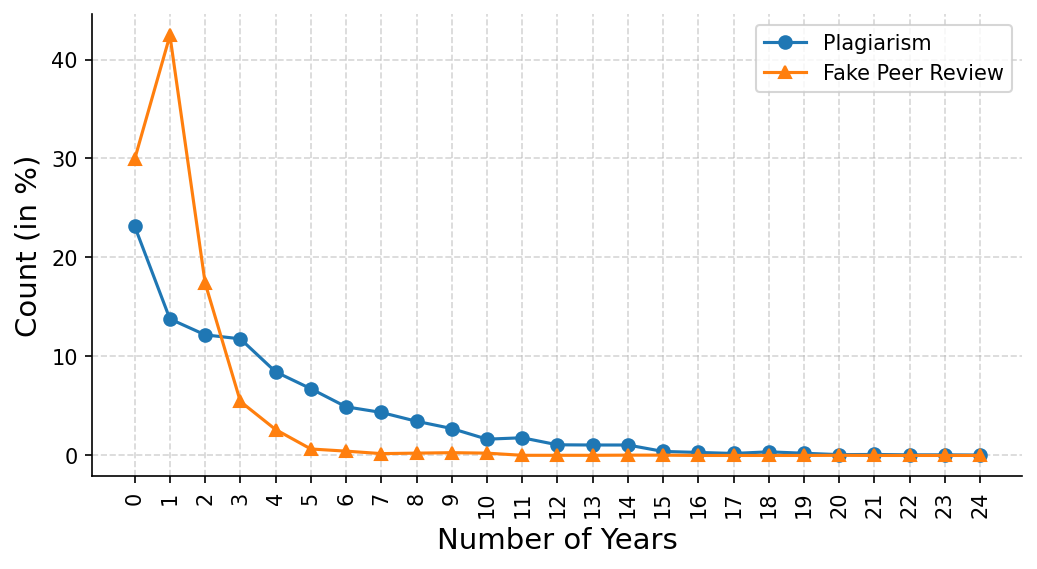} 

\caption{Retraction time.}
\label{fig:RetTime}
\end{figure}

Figure~\ref{fig:RetTime} visualizes the retraction trends over a period of years for two categories: plagiarism and fake peer review. Plagiarism starts at a high percentage (around 23.2\%) for papers retracted within the first year of publication. It gradually decreases over the years, showing that plagiarism cases are identified relatively early. However, fake peer review increases sharply in the first year, reaching over 42.4\%, indicating that these cases are caught quickly after publication. The percentage drops steeply within the first few years and approaches zero after about 5 years, implying that this issue is typically resolved early. Overall, the fake peer review cases are identified and retracted more rapidly than plagiarism cases.
%~~~~~~~~~~~~~~~~~~~~~~~~~~~~~~~~~~~~~~~~~~~~~~~~~~~~~~~~~~~~~~~~~~~~~
%~~~~~~~~~~~~~~~~~~~~~~~~~~~~~~~~~~~~~~~~~~~~~~~~~~~~~~~~~~~~~~~~~~~~~~~

%==================================================================
\subsection{Analysis of Retracted Citations}

In 2009, the Committee on Publication Ethics (COPE)~\citep{barbourguidelines} released retraction guidelines recommending that notices clearly explain the reasons for retraction and distinguish misconduct from honest error. These notices should be freely accessible and linked to the retracted article to prevent unintentional citations. A study of MEDLINE retracted articles (1966–1997) found that 94\% of the citations to retracted works were made unknowingly~\citep{budd1998phenomena}. \cite{DeCassai2022Inappropriate} also highlighted in their study that 89\% of the authors were unaware that they cited retracted articles which may be due to inadequate notification in journals and stored copies.

The results in Table \ref{tab:RetCitations} and Table \ref{tab:CitationCategory} provide an overview of retracted and non-retracted citations across two retraction categories - Plagiarism and Fake Peer Review - and offer a detailed breakdown of retraction categories for retracted citations. Plagiarism accounts for a significantly higher total number of citations compared to fake peer review. Plagiarism has 98.4\% non-retracted citations and 1.6\% retracted citations, with 46.1\% of its retracted citations attributed to plagiarism itself.
Fake peer review has 97.6\% non-retracted citations and 2.4\% retracted citations, with 53.6\% of its retracted citations attributed to fake peer review.
%~~~~~~~~~~~~~~~~~~~~~~~~~~~~~~~~~~~~~~~~~~~~~~~~~~~~~~~~~~~~~~~~~~~~~~~
\begin{table}[H]
\caption{Number of retracted and non-retracted citations in both categories.}
\begin{tabular}{|l|c|cc|c|}
\hline
\multicolumn{1}{|c|}{\multirow{2}{*}{\textbf{Retraction Category}}} & \multirow{2}{*}{\textbf{Total Citations}} & \multicolumn{2}{c|}{\textbf{Number of Citations}}               & \multirow{2}{*}{\textbf{\begin{tabular}[c]{@{}c@{}}Mapped with \\ Retraction Watch\end{tabular}}} \\ \cline{3-4}
\multicolumn{1}{|c|}{}                                              &                                           & \multicolumn{1}{c|}{\textbf{NonRetracted}} & \textbf{Retracted} &                                                                                                   \\ \hline
Plagiarism                                                          & 1,41,891                                    & \multicolumn{1}{c|}{1,39,621}                & 2,270               & 2,100                                                                                              \\ \hline
Fake Peer Review                                                    & 55,272                                     & \multicolumn{1}{c|}{53,929}                 & 1,343               & 1,138                                                                                              \\ \hline
Total                                                               & 1,97,163                                    & \multicolumn{1}{c|}{1,93,550}                & 3,613               & 3,238                                                                                              \\ \hline
\end{tabular}
\label{tab:RetCitations}
\end{table}
%~~~~~~~~~~~~~~~~~~~~~~~~~~~~~~~~~~~~~~~~~~~~~~~~~~~~~~~~~~~~~~~~~~~~~~~
%~~~~~~~~~~~~~~~~~~~~~~~~~~~~~~~~~~~~~~~~~~~~~~~~~~~~~~~~~~~~~~~~~~~~~~~
\begin{table}[H]
\centering
\caption{Retraction category of retracted citations.}
\begin{tabular}{|l|ccc|c|}
\hline
\multirow{2}{*}{\textbf{Retraction Category}} & \multicolumn{3}{c|}{\textbf{Retraction Category}}                                                           & \multirow{2}{*}{\textbf{\begin{tabular}[c]{@{}c@{}}Total Retracted \\ Citations\end{tabular}}} \\ \cline{2-4}& \multicolumn{1}{c|}{\textbf{Plagiarism}} & \multicolumn{1}{c|}{\textbf{Fake Peer Review}} & \textbf{Others} &                                                                                                \\ \hline
Plagiarism                                    & \multicolumn{1}{c|}{967}                 & \multicolumn{1}{c|}{76}                        & 1,057            & 2,100                                                                                           \\ \hline
Fake Peer Review                              & \multicolumn{1}{c|}{49}                  & \multicolumn{1}{c|}{610}                       & 479             & 1,138                                                                                           \\ \hline
Total                                         & \multicolumn{1}{c|}{1,016}                & \multicolumn{1}{c|}{686}                       & 1,536            & 3,238                                                                                           \\ \hline
\end{tabular}
\label{tab:CitationCategory}
\end{table}
%~~~~~~~~~~~~~~~~~~~~~~~~~~~~~~~~~~~~~~~~~~~~~~~~~~~~~~~~~~~~~~~~~~~~~~~
%==================================================================
\subsection{Analysis of Self-citations}

Self-citation in research refers to the practice in which authors cite their own previous works in new publications. Self-citations significantly contribute to the continued citation of retracted articles, and approximately 18\% of the authors cite their own retracted work after retraction. There is also a positive correlation between self-citations and the total number of citations after retraction~\citep{Madlock-Brown2014The}. 
After analyzing Table~\ref{tab:RetCitations} and Table~\ref{tab:SelfCite}, it is observed that only 1.49\% of citations to retracted papers are self-citations in cases of plagiarism, and 1.96\% in cases of fake peer review. Furthermore, when examining citations that were later retracted, 17.18\% of these are self-citations in cases of plagiarism and 13\% in cases of fake peer review. 
Overall, plagiarism involves fewer self-citations compared to fake peer review but has a higher proportion of retracted self-citations.
%~~~~~~~~~~~~~~~~~~~~~~~~~~~~~~~~~~~~~~~~~~~~~~~~~~~~~~~~~~~~~~~~~~~~~~~
\begin{table}[!ht]
\centering
\caption{Number of self-citations to retracted papers. A total of 3,205 self-citations are subset of 11,344 retracted papers.}
\begin{tabular}{|l|ccc|}
\hline
\multirow{2}{*}{\textbf{Retraction Category}} & \multicolumn{3}{c|}{\textbf{Self-citations}} \\ \cline{2-4}& \multicolumn{1}{c|}{\textbf{Total}} & \multicolumn{1}{c|}{\textbf{Retracted}} & \textbf{Non Retracted} \\ \hline
Plagiarism                                    & \multicolumn{1}{c|}{2119}           & \multicolumn{1}{c|}{390}                & 1729                   \\ \hline
Fake Peer Review                              & \multicolumn{1}{c|}{1086}           & \multicolumn{1}{c|}{175}                & 911                    \\ \hline
Total                                         & \multicolumn{1}{c|}{3205}           & \multicolumn{1}{c|}{565}                & 2640                   \\ \hline
\end{tabular}
\label{tab:SelfCite}
\end{table}
%~~~~~~~~~~~~~~~~~~~~~~~~~~~~~~~~~~~~~~~~~~~~~~~~~~~~~~~~~~~~~~~~~~~~~~~

Furthermore, Table~\ref{tab:TeamSize} presents the distribution of author pairs who self-cited their retracted papers. As shown in Table~\ref{tab:CitationCategory}, a total of 11,344 retracted papers received 197,163 citations. Of these, 8.5\% were self-citations. Among the self-citations, individual authors (including repetitions) contributed 4.18\%, with 70\% of these citations directed toward papers categorized as plagiarized. Similarly, 2.13\% of the citations were self-citations by pairs of authors (teams consisting of two authors, including repetitions but limited to groups of two). Within these 2.13\% self-citations, most (76.79\%) were citations to articles classified under the plagiarized category. In general, 75.4\% of self-citations from different groups of authors were associated with plagiarized articles. This count of self-citations under plagiarized category keeps on increasing as the team size increases.

%~~~~~~~~~~~~~~~~~~~~~~~~~~~~~~~~~~~~~~~~~~~~~~~~~~~~~~~~~~~~~~~~~~~~~~~
\begin{table}[!ht]
\centering
\caption{Team size of authors who self-cited their retracted papers. A total of 16,871 self-citations is a subset of 197,163 citations received by 11,344 retracted papers.}
\begin{tabular}{|c|ccc|l|c|ccc|}
\hline
\multirow{2}{*}{\textbf{Team Size}} & \multicolumn{3}{c|}{\textbf{Self-citations}} &  & \multirow{2}{*}{\textbf{Team Size}} & \multicolumn{3}{c|}{\textbf{Self-citations}} \\ \cline{2-4} \cline{7-9}  & \multicolumn{1}{l|}{\textbf{Total}} & \multicolumn{1}{c|}{\textbf{Plagiarism}} & \textbf{Fake Peer Review} &  & & \multicolumn{1}{l|}{\textbf{Total}} & \multicolumn{1}{c|}{\textbf{Plagiarism}} & \textbf{Fake Peer Review} \\ \cline{1-4} \cline{6-9} 

1                                   & \multicolumn{1}{c|}{8248}           & \multicolumn{1}{c|}{5778}                & 2470                      &  & 10                                  & \multicolumn{1}{c|}{37}             & \multicolumn{1}{c|}{35}                  & 2                         \\ \cline{1-4} \cline{6-9} 
2                                   & \multicolumn{1}{c|}{4201}           & \multicolumn{1}{c|}{3226}                & 975                       &  & 11                                  & \multicolumn{1}{c|}{23}             & \multicolumn{1}{c|}{23}                  & -                         \\ \cline{1-4} \cline{6-9} 
3                                   & \multicolumn{1}{c|}{2084}           & \multicolumn{1}{c|}{1697}                & 387                       &  & 12                                  & \multicolumn{1}{c|}{4}              & \multicolumn{1}{c|}{4}                   & -                         \\ \cline{1-4} \cline{6-9} 
4                                   & \multicolumn{1}{c|}{1080}           & \multicolumn{1}{c|}{914}                 & 166                       &  & 13                                  & \multicolumn{1}{c|}{4}              & \multicolumn{1}{c|}{4}                   & -                         \\ \cline{1-4} \cline{6-9} 
5                                   & \multicolumn{1}{c|}{608}            & \multicolumn{1}{c|}{516}                 & 92                        &  & 14                                  & \multicolumn{1}{c|}{6}              & \multicolumn{1}{c|}{5}                   & 1                         \\ \cline{1-4} \cline{6-9} 
6                                   & \multicolumn{1}{c|}{262}            & \multicolumn{1}{c|}{237}                 & 25                        &  & 15                                  & \multicolumn{1}{c|}{4}              & \multicolumn{1}{c|}{4}                   & -                         \\ \cline{1-4} \cline{6-9} 
7                                   & \multicolumn{1}{c|}{167}            & \multicolumn{1}{c|}{152}                 & 15                        &  & 17                                  & \multicolumn{1}{c|}{1}              & \multicolumn{1}{c|}{1}                   & -                         \\ \cline{1-4} \cline{6-9} 
8                                   & \multicolumn{1}{c|}{91}             & \multicolumn{1}{c|}{81}                  & 10                        &  & 19                                  & \multicolumn{1}{c|}{1}              & \multicolumn{1}{c|}{1}                   & -                         \\ \cline{1-4} \cline{6-9} 
9                                   & \multicolumn{1}{c|}{50}             & \multicolumn{1}{c|}{47}                  & 3    &  & Total                               & \multicolumn{1}{c|}{16871}          & \multicolumn{1}{c|}{12725}               & 4146                      \\ \hline
\end{tabular}
\label{tab:TeamSize}
\end{table}
%==================================================================
\section{Conclusion}
Citing retracted studies is an important issue in academia because it risks spreading misinformation and undermining the integrity of the scientific literature~\citep{van2024exclusive}. Although retraction notices are issued, many retracted papers are still cited without noting their retracted status. This problem occurs in various fields, including computer science and biomedical research, where retracted papers are often cited in systematic reviews and meta-analyses. A major cause for this is that the authors are not informed about the status of article retractions, either because they do not receive enough notifications in journals and databases or because they depend on saved copies or find uncorrected versions available on open-access platforms~\citep{Million2024Disinformation}.

The management and identification of retracted publications are challenging due to logistical issues and the decentralized nature of publication databases. Retraction notices are often not prominently displayed, and databases frequently fail to effectively link retracted articles to their corresponding notices. Improved visibility of retraction notices is essential, including clear labeling and alerts to prevent the continued citation of retracted articles. The authors must verify with diligence the status of the retraction of the articles they cite, and the reviewers must check that references in the manuscripts are current and correct.

The continued existence of citations to retracted articles, with a role played by self-citations, requires improved retraction practices and better awareness among scientists. An increased awareness of the consequences of citing retracted work, in conjunction with providing education on how to check the status of the article, can significantly reduce inappropriate citations~\citep{DeCassai2023Citing, minetto2023p}. This will help ensure that the scientific literature remains credible and of high quality.
%==================================================================
%==================================================================
\section*{Data Availability Statement}
The datasets used in the study will be available from the corresponding author on request.

\section*{Acknowledgment}
Authors gratefully acknowledges the Research and Development Cell, BML Munjal University for
their financial support through the seed grant (No: BMU/RDC/SG/2024-06), which made this research possible.

 \section*{Conflict of interest}
 The authors declare no conflict of interest.

%%==============================================
\bibliographystyle{cas-model2-names}
\bibliography{Reference}

\begin{thebibliography}{33}
\expandafter\ifx\csname natexlab\endcsname\relax\def\natexlab#1{#1}\fi
\providecommand{\url}[1]{\texttt{#1}}
\providecommand{\href}[2]{#2}
\providecommand{\path}[1]{#1}
\providecommand{\DOIprefix}{doi:}
\providecommand{\ArXivprefix}{arXiv:}
\providecommand{\URLprefix}{URL: }
\providecommand{\Pubmedprefix}{pmid:}
\providecommand{\doi}[1]{\href{http://dx.doi.org/#1}{\path{#1}}}
\providecommand{\Pubmed}[1]{\href{pmid:#1}{\path{#1}}}
\providecommand{\bibinfo}[2]{#2}
\ifx\xfnm\relax \def\xfnm[#1]{\unskip,\space#1}\fi
%Type = Article
\bibitem[{Bar-Ilan and Halevi(2017)}]{Bar-Ilan2017Post}
\bibinfo{author}{Bar-Ilan, J.}, \bibinfo{author}{Halevi, G.}, \bibinfo{year}{2017}.
\newblock \bibinfo{title}{Post retraction citations in context: a case study}.
\newblock \bibinfo{journal}{Scientometrics} \bibinfo{volume}{113}, \bibinfo{pages}{547 -- 565}.
\newblock \DOIprefix\doi{10.1007/s11192-017-2242-0}.
%Type = Misc
\bibitem[{Barbour et~al.(2009)Barbour, Kleinert, Wager and Yentis}]{barbourguidelines}
\bibinfo{author}{Barbour, V.}, \bibinfo{author}{Kleinert, S.}, \bibinfo{author}{Wager, E.}, \bibinfo{author}{Yentis, S.}, \bibinfo{year}{2009}.
\newblock \bibinfo{title}{Guidelines for retracting articles. committee on publication ethics; 2009 sep}.
\newblock \DOIprefix\doi{https://doi.org/10.24318/cope.2019.1.4}.
%Type = Article
\bibitem[{Bell et~al.(2022)Bell, Kingori and Mills}]{Bell2022Scholarly}
\bibinfo{author}{Bell, K.}, \bibinfo{author}{Kingori, P.}, \bibinfo{author}{Mills, D.S.}, \bibinfo{year}{2022}.
\newblock \bibinfo{title}{Scholarly publishing, boundary processes, and the problem of fake peer reviews}.
\newblock \bibinfo{journal}{Science, technology \& human values} \bibinfo{volume}{49}, \bibinfo{pages}{78 -- 104}.
\newblock \DOIprefix\doi{10.1177/01622439221112463}.
%Type = Article
\bibitem[{Bolboacă et~al.(2019)Bolboacă, Buhai, Aluaș and Bulboacă}]{Bolboaca2019Post}
\bibinfo{author}{Bolboacă, S.D.}, \bibinfo{author}{Buhai, D.}, \bibinfo{author}{Aluaș, M.}, \bibinfo{author}{Bulboacă, A.}, \bibinfo{year}{2019}.
\newblock \bibinfo{title}{Post retraction citations among manuscripts reporting a radiology-imaging diagnostic method}.
\newblock \bibinfo{journal}{PLoS ONE} \bibinfo{volume}{14}.
\newblock \DOIprefix\doi{10.1371/journal.pone.0217918}.
%Type = Article
\bibitem[{Budd et~al.(1998)Budd, Sievert and Schultz}]{budd1998phenomena}
\bibinfo{author}{Budd, J.M.}, \bibinfo{author}{Sievert, M.}, \bibinfo{author}{Schultz, T.R.}, \bibinfo{year}{1998}.
\newblock \bibinfo{title}{Phenomena of retraction: reasons for retraction and citations to the publications}.
\newblock \bibinfo{journal}{Jama} \bibinfo{volume}{280}, \bibinfo{pages}{296--297}.
%Type = Article
\bibitem[{Cassai et~al.(2022)Cassai, Geraldini, Pinto, Carbonari, Cascella, Boscolo, Sella, Monteleone, Cavaliere, Munari, Garofalo and Navalesi}]{DeCassai2022Inappropriate}
\bibinfo{author}{Cassai, A.D.}, \bibinfo{author}{Geraldini, F.}, \bibinfo{author}{Pinto, S.D.}, \bibinfo{author}{Carbonari, I.}, \bibinfo{author}{Cascella, M.}, \bibinfo{author}{Boscolo, A.}, \bibinfo{author}{Sella, N.}, \bibinfo{author}{Monteleone, F.}, \bibinfo{author}{Cavaliere, F.}, \bibinfo{author}{Munari, M.}, \bibinfo{author}{Garofalo, E.}, \bibinfo{author}{Navalesi, P.}, \bibinfo{year}{2022}.
\newblock \bibinfo{title}{Inappropriate citation of retracted articles in anesthesiology and intensive care medicine publications}.
\newblock \bibinfo{journal}{Anesthesiology} \bibinfo{volume}{137}, \bibinfo{pages}{341 -- 350}.
\newblock \DOIprefix\doi{10.1097/ALN.0000000000004302}.
%Type = Article
\bibitem[{Cassai et~al.(2023)Cassai, Volpe, Geraldini, Dost, Boscolo and Navalesi}]{DeCassai2023Citing}
\bibinfo{author}{Cassai, A.D.}, \bibinfo{author}{Volpe, F.}, \bibinfo{author}{Geraldini, F.}, \bibinfo{author}{Dost, B.}, \bibinfo{author}{Boscolo, A.}, \bibinfo{author}{Navalesi, P.}, \bibinfo{year}{2023}.
\newblock \bibinfo{title}{Citing retracted literature: a word of caution}.
\newblock \bibinfo{journal}{Regional Anesthesia \& Pain Medicine} \bibinfo{volume}{48}, \bibinfo{pages}{349 -- 351}.
\newblock \DOIprefix\doi{10.1136/rapm-2022-104177}.
%Type = Article
\bibitem[{Gray et~al.(2019)Gray, Al-Ghareeb and Mckenna}]{Gray2019Why}
\bibinfo{author}{Gray, R.}, \bibinfo{author}{Al-Ghareeb, A.}, \bibinfo{author}{Mckenna, L.}, \bibinfo{year}{2019}.
\newblock \bibinfo{title}{Why articles continue to be cited after they have been retracted: An audit of retraction notices.}
\newblock \bibinfo{journal}{International journal of nursing studies} \bibinfo{volume}{90}, \bibinfo{pages}{11--12}.
\newblock \DOIprefix\doi{10.1016/j.ijnurstu.2018.10.003}.
%Type = Article
\bibitem[{Hadi(2016)}]{Hadi2016Fake}
\bibinfo{author}{Hadi, M.A.}, \bibinfo{year}{2016}.
\newblock \bibinfo{title}{Fake peer‐review in research publication: revisiting research purpose and academic integrity}.
\newblock \bibinfo{journal}{International Journal of Pharmacy Practice} \bibinfo{volume}{24}.
\newblock \DOIprefix\doi{10.1111/ijpp.12307}.
%Type = Article
\bibitem[{Hagberg(2020)}]{Hagberg2020The}
\bibinfo{author}{Hagberg, J.}, \bibinfo{year}{2020}.
\newblock \bibinfo{title}{The unfortunately long life of some retracted biomedical research publications.}
\newblock \bibinfo{journal}{Journal of applied physiology} \DOIprefix\doi{10.1152/japplphysiol.00003.2020}.
%Type = Article
\bibitem[{Hamilton(2019)}]{hamilton2019continued}
\bibinfo{author}{Hamilton, D.G.}, \bibinfo{year}{2019}.
\newblock \bibinfo{title}{Continued citation of retracted radiation oncology literature—do we have a problem?}
\newblock \bibinfo{journal}{International Journal of Radiation Oncology* Biology* Physics} \bibinfo{volume}{103}, \bibinfo{pages}{1036--1042}.
%Type = Article
\bibitem[{Heibi and Peroni(2021)}]{heibi2021qualitative}
\bibinfo{author}{Heibi, I.}, \bibinfo{author}{Peroni, S.}, \bibinfo{year}{2021}.
\newblock \bibinfo{title}{A qualitative and quantitative analysis of open citations to retracted articles: the wakefield 1998 et al.'s case}.
\newblock \bibinfo{journal}{Scientometrics} \bibinfo{volume}{126}, \bibinfo{pages}{8433--8470}.
%Type = Article
\bibitem[{Heibi and Peroni(2022)}]{heibi2022protocol}
\bibinfo{author}{Heibi, I.}, \bibinfo{author}{Peroni, S.}, \bibinfo{year}{2022}.
\newblock \bibinfo{title}{A protocol to gather, characterize and analyze incoming citations of retracted articles}.
\newblock \bibinfo{journal}{Plos one} \bibinfo{volume}{17}, \bibinfo{pages}{e0270872}.
%Type = Article
\bibitem[{Kamali et~al.(2020)Kamali, abadi and Rahimi}]{Kamali2020Plagiarism}
\bibinfo{author}{Kamali, N.}, \bibinfo{author}{abadi, A.T.B.}, \bibinfo{author}{Rahimi, F.}, \bibinfo{year}{2020}.
\newblock \bibinfo{title}{Plagiarism, fake peer-review, and duplication: Predominant reasons underlying retractions of iran-affiliated scientific papers}.
\newblock \bibinfo{journal}{Science and Engineering Ethics} \bibinfo{volume}{26}, \bibinfo{pages}{3455 -- 3463}.
\newblock \DOIprefix\doi{10.1007/s11948-020-00274-6}.
%Type = Inproceedings
\bibitem[{Khurana et~al.(2022)Khurana, Ganesan, Kumar and Sharma}]{khurana2022bibliometric}
\bibinfo{author}{Khurana, P.}, \bibinfo{author}{Ganesan, G.}, \bibinfo{author}{Kumar, G.}, \bibinfo{author}{Sharma, K.}, \bibinfo{year}{2022}.
\newblock \bibinfo{title}{A bibliometric analysis to unveil the impact of digital object identifiers (doi) on bibliometric indicators}, in: \bibinfo{booktitle}{Proceedings of Third International Conference on Computing, Communications, and Cyber-Security: IC4S 2021}, \bibinfo{organization}{Springer}. pp. \bibinfo{pages}{859--869}.
%Type = Article
\bibitem[{Koçyiğit et~al.(2023)Koçyiğit, Akyol, Zhaksylyk, Seiil and Yessirkepov}]{Kocyigit2023Analysis}
\bibinfo{author}{Koçyiğit, B.}, \bibinfo{author}{Akyol, A.}, \bibinfo{author}{Zhaksylyk, A.}, \bibinfo{author}{Seiil, B.}, \bibinfo{author}{Yessirkepov, M.}, \bibinfo{year}{2023}.
\newblock \bibinfo{title}{Analysis of retracted publications in medical literature due to ethical violations}.
\newblock \bibinfo{journal}{Journal of Korean Medical Science} \bibinfo{volume}{38}.
\newblock \DOIprefix\doi{10.3346/jkms.2023.38.e324}.
%Type = Article
\bibitem[{Kühberger et~al.(2022)Kühberger, Streit and Scherndl}]{Kuhberger2022Self-correction}
\bibinfo{author}{Kühberger, A.}, \bibinfo{author}{Streit, D.}, \bibinfo{author}{Scherndl, T.}, \bibinfo{year}{2022}.
\newblock \bibinfo{title}{Self-correction in science: The effect of retraction on the frequency of citations}.
\newblock \bibinfo{journal}{PLOS ONE} \bibinfo{volume}{17}.
\newblock \DOIprefix\doi{10.1371/journal.pone.0277814}.
%Type = Article
\bibitem[{Madlock-Brown and Eichmann(2014)}]{Madlock-Brown2014The}
\bibinfo{author}{Madlock-Brown, C.}, \bibinfo{author}{Eichmann, D.}, \bibinfo{year}{2014}.
\newblock \bibinfo{title}{The (lack of) impact of retraction on citation networks}.
\newblock \bibinfo{journal}{Science and Engineering Ethics} \bibinfo{volume}{21}, \bibinfo{pages}{127 -- 137}.
\newblock \DOIprefix\doi{10.1007/s11948-014-9532-1}.
%Type = Article
\bibitem[{Million and Budd(2024)}]{Million2024Disinformation}
\bibinfo{author}{Million, A.J.}, \bibinfo{author}{Budd, J.M.}, \bibinfo{year}{2024}.
\newblock \bibinfo{title}{Disinformation in science: Ethical considerations for citing retracted works}.
\newblock \bibinfo{journal}{Proceedings of the Association for Information Science and Technology} \DOIprefix\doi{10.1002/pra2.1025}.
%Type = Article
\bibitem[{Minetto et~al.(2023)Minetto, Pisaturo, Cermisoni, Rabellotti, Pagliardini, Candiani, Papaleo and Alteri}]{minetto2023p}
\bibinfo{author}{Minetto, S.}, \bibinfo{author}{Pisaturo, D.}, \bibinfo{author}{Cermisoni, G.}, \bibinfo{author}{Rabellotti, E.}, \bibinfo{author}{Pagliardini, L.}, \bibinfo{author}{Candiani, M.}, \bibinfo{author}{Papaleo, E.}, \bibinfo{author}{Alteri, A.}, \bibinfo{year}{2023}.
\newblock \bibinfo{title}{P-385 are you completely aware of your citations? a cross-sectional study on improper citations of retracted articles in medically assisted reproduction}.
\newblock \bibinfo{journal}{Human Reproduction} \bibinfo{volume}{38}, \bibinfo{pages}{dead093--349}.
%Type = Article
\bibitem[{Palla et~al.(2023)Palla, Singson and Thiyagarajan}]{Palla2023Systematic}
\bibinfo{author}{Palla, I.A.}, \bibinfo{author}{Singson, M.}, \bibinfo{author}{Thiyagarajan, S.}, \bibinfo{year}{2023}.
\newblock \bibinfo{title}{Systematic examination of post‐ and pre‐citation of indian‐authored retracted papers}.
\newblock \bibinfo{journal}{Learned Publishing} \bibinfo{volume}{36}.
\newblock \DOIprefix\doi{10.1002/leap.1572}.
%Type = Article
\bibitem[{Qi et~al.(2016)Qi, Deng and Guo}]{Qi2016Characteristics}
\bibinfo{author}{Qi, X.}, \bibinfo{author}{Deng, H.}, \bibinfo{author}{Guo, X.}, \bibinfo{year}{2016}.
\newblock \bibinfo{title}{Characteristics of retractions related to faked peer reviews: an overview}.
\newblock \bibinfo{journal}{Postgraduate Medical Journal} \bibinfo{volume}{93}, \bibinfo{pages}{499 -- 503}.
\newblock \DOIprefix\doi{10.1136/postgradmedj-2016-133969}.
%Type = Article
\bibitem[{Rivera(2018)}]{Rivera2018Fake}
\bibinfo{author}{Rivera, H.}, \bibinfo{year}{2018}.
\newblock \bibinfo{title}{Fake peer review and inappropriate authorship are real evils}.
\newblock \bibinfo{journal}{Journal of Korean Medical Science} \bibinfo{volume}{34}.
\newblock \DOIprefix\doi{10.3346/jkms.2019.34.e6}.
%Type = Article
\bibitem[{Schneider et~al.(2020)Schneider, Ye, Ye, Hill, Whitehorn and Whitehorn}]{Schneider2020Continued}
\bibinfo{author}{Schneider, J.}, \bibinfo{author}{Ye, D.}, \bibinfo{author}{Ye, D.}, \bibinfo{author}{Hill, A.M.}, \bibinfo{author}{Whitehorn, A.}, \bibinfo{author}{Whitehorn, A.}, \bibinfo{year}{2020}.
\newblock \bibinfo{title}{Continued post-retraction citation of a fraudulent clinical trial report, 11 years after it was retracted for falsifying data}.
\newblock \bibinfo{journal}{Scientometrics} \bibinfo{volume}{125}, \bibinfo{pages}{2877 -- 2913}.
\newblock \DOIprefix\doi{10.1007/s11192-020-03631-1}.
%Type = Article
\bibitem[{Sharma(2021)}]{sharma2021team}
\bibinfo{author}{Sharma, K.}, \bibinfo{year}{2021}.
\newblock \bibinfo{title}{Team size and retracted citations reveal the patterns of retractions from 1981 to 2020}.
\newblock \bibinfo{journal}{Scientometrics} \bibinfo{volume}{126}, \bibinfo{pages}{8363--8374}.
%Type = Article
\bibitem[{Sharma(2024)}]{sharma2024over}
\bibinfo{author}{Sharma, K.}, \bibinfo{year}{2024}.
\newblock \bibinfo{title}{Over two decades of scientific misconduct in india: Retraction reasons and journal quality among inter-country and intra-country institutional collaboration}.
\newblock \bibinfo{journal}{Scientometrics} , \bibinfo{pages}{1--23}.
%Type = Article
\bibitem[{da~Silva(2020)}]{DaSilva2020Reasons}
\bibinfo{author}{da~Silva, J.A.T.}, \bibinfo{year}{2020}.
\newblock \bibinfo{title}{Reasons for citing retracted literature are not straightforward, and solutions are complex.}
\newblock \bibinfo{journal}{Journal of applied physiology} \bibinfo{volume}{129 1}, \bibinfo{pages}{3}.
\newblock \DOIprefix\doi{10.1152/japplphysiol.00258.2020}.
%Type = Article
\bibitem[{Silva and Bornemann-Cimenti(2016)}]{Silva2016Why}
\bibinfo{author}{Silva, J.A.T.}, \bibinfo{author}{Bornemann-Cimenti, H.}, \bibinfo{year}{2016}.
\newblock \bibinfo{title}{Why do some retracted papers continue to be cited?}
\newblock \bibinfo{journal}{Scientometrics} \bibinfo{volume}{110}, \bibinfo{pages}{365--370}.
\newblock \DOIprefix\doi{10.1007/s11192-016-2178-9}.
%Type = Article
\bibitem[{Steen et~al.(2013)Steen, Casadevall and Fang}]{Steen2013Why}
\bibinfo{author}{Steen, R.}, \bibinfo{author}{Casadevall, A.}, \bibinfo{author}{Fang, F.}, \bibinfo{year}{2013}.
\newblock \bibinfo{title}{Why has the number of scientific retractions increased?}
\newblock \bibinfo{journal}{PLoS ONE} \bibinfo{volume}{8}.
\newblock \DOIprefix\doi{10.1371/journal.pone.0068397}.
%Type = Article
\bibitem[{Tang(2023)}]{Tang2023Some}
\bibinfo{author}{Tang, B.}, \bibinfo{year}{2023}.
\newblock \bibinfo{title}{Some insights into the factors influencing continuous citation of retracted scientific papers}.
\newblock \bibinfo{journal}{Publ.} \bibinfo{volume}{11}, \bibinfo{pages}{47}.
\newblock \DOIprefix\doi{10.3390/publications11040047}.
%Type = Article
\bibitem[{Van~Noorden and Naddaf(2024)}]{van2024exclusive}
\bibinfo{author}{Van~Noorden, R.}, \bibinfo{author}{Naddaf, M.}, \bibinfo{year}{2024}.
\newblock \bibinfo{title}{Exclusive: the papers that most heavily cite retracted studies}.
\newblock \bibinfo{journal}{Nature} \bibinfo{volume}{633}, \bibinfo{pages}{13--15}.
%Type = Article
\bibitem[{Wang and Chen(2025)}]{wang2025empirical}
\bibinfo{author}{Wang, D.}, \bibinfo{author}{Chen, S.}, \bibinfo{year}{2025}.
\newblock \bibinfo{title}{An empirical study of retractions due to honest errors: Exploring the relationship between error types and author teams}.
\newblock \bibinfo{journal}{Journal of Informetrics} \bibinfo{volume}{19}, \bibinfo{pages}{101600}.
%Type = Article
\bibitem[{Wang and Su(2022)}]{Wang2022Expert-recommended}
\bibinfo{author}{Wang, P.}, \bibinfo{author}{Su, J.}, \bibinfo{year}{2022}.
\newblock \bibinfo{title}{Expert-recommended biomedical journal articles: Their retractions or corrections, and post-retraction citing}.
\newblock \bibinfo{journal}{Journal of Information Science} \bibinfo{volume}{50}, \bibinfo{pages}{17 -- 34}.
\newblock \DOIprefix\doi{10.1177/01655515221074329}.

\end{thebibliography}
%==================================================================
\end{document}